# Paper-based spintronics: magneto-resistivity of permalloy deposited onto paper substrates

Meriem Akin, Lutz Rissing
Institute of Micro Production Technology
An der Universitaet 2, 30823 Garbsen, Germany
akin@impt.uni-hannover.de

**Abstract**

Driven by low-cost, resource abundance and distinct material properties, the use of paper in electronics, optics and fluidics is under investigation. In order to realize a dense coverage of sensor networks as part of the roadmap of the internet-of-things, achieving lower manufacturing cost of the aforementioned sensors is required. Considering sensor systems based on magneto-resistance principles (anisotropic, giant, tunnel) that are conventionally manufactured onto inorganic semiconductor materials, we propose the use of paper substrates for cost reduction purposes primarily. In particular, we studied the magneto-resistance sensitivity of permalloy (Py:$Ni_{81}Fe_{19}$) onto paper substrates. In this work, we report on our findings with clean room paper (80 g/m², $R_{rms}$ = 2.877 µm, 23% surface porosity, latex impregnation, no embossed macro-structure). Here, the Py:$Ni_{81}Fe_{19}$ coating was manufactured by means of a dry process, sputter deposition, and spans an area of 10x10 mm² and a thickness of 70 nm. Employing a four-point-probe DC resistivity measurement setup, we investigated the change of electrical resistance of Py:$Ni_{81}Fe_{19}$ under the presence of an oriented external magnetic field. In particular, we investigate the magneto-resistive change at two configurations: (1) the direction of the magnetic field is parallel to the nominal induced electric current and (2) the direction of the magnetic field is perpendicular to the electric current. Due to the stochastic orientation of the fibers interplaying with the Py:$Ni_{81}Fe_{19}$ coating, the change in magneto-resistance of the overall system at both measurement configurations closely corresponds to the classical response of Py:$Ni_{81}Fe_{19}$ at a ±45° angle between the direction of electrical current and magnetic field. Using the magneto-optic kerr effect, we observed the formation of domain walls at the fiber bending locations. Future work will focus on the impact of layer thickness, fiber dimensions and structure of magnetic coating on the performance of the paper-based Py:$Ni_{81}Fe_{19}$ magneto-resistors.

**Introduction**

With the digitization of information, paper is moving away from its classical fields of application such as writing and printing of text. Yet, today, paper still exists despite the belief of many visionaries over the centuries [1]. In fact, several disruptive applications of paper have erupted such as the usage of paper tubes in construction of buildings [2], interactive paper newsprints with embedded circuits [3], diagnosis micro-labs made of paper [4], paper-based foldable microscopes [5], etc. [6]. In terms of cost analysis, the average cost of 1 m² of variously engineered paper ranges from [0.10, 2.00] US$ while the cost of 1 m² of widely used polyimide films and 1 m² of polished silicon are about 5.00 US$ and 1,000 US$ as of 2013 respectively[1]. In addition to its low cost, paper is an abundant resource with a wide base of qualifying technical properties (biocompatibility, chemical stability, mechanical bendability etc.). Thus, while paper is being reinvented to a smart physical object with embedded sensors and software, it is compellingly becoming part of the evolving network of the internet-of-things.

In order to achieve high-yield and high-performing systems on paper, various efforts on characterizing paper at macro-, micro- and nanoscales abide ([7], [8], [9]). Besides, new paper types are being fabricated to conform to specialized engineering applications. For instance, a transparent smooth paper (>90% optical transmission and 10 nm surface roughness) was developed and demonstrated in manufacturing opto-electronic devices [10]. By employing vulcanized paper fibers, strong paper was fabricated that was shown to work for mechanically stressed parts in the automotive industry [11]. Also, by multi-layering paper fibers with conductive polymers [12] or growing conductive particles onto paper fibers [13], paper was converted from an insulating carrier to a functional conductor.

In this work, we are interested in employing paper in magnetic based micro-electro-mechanical systems for the aforementioned benefits. In particular, we would like to explore the realization of various magneto-resistance effects such as anisotropic magneto-resistance, giant magneto-resistance, tunnel magneto-resistance etc. onto paper platforms. Hence, we would like to investigate the precision, i.e. uniqueness of the magnetic response on paper. We envision paper to make these specialized and mostly expensive spintronic systems such as spin-valve sensors, magnetic compasses and read/write heads available at a much lower cost for a broader user base. Thus, we study the interplay between the physical properties of paper and the magneto-electric phenomena. Besides, we investigate manufacturability and propose fabrication processes for mass-production purposes.

**Prior art on magnetics in paper-based systems**

Magnetism as an added value to the existing technical properties of paper has been put for years into practice e.g. in the thick magnetic stripes laminated onto subway and parking garage tickets. Also, by chemically synthesizing magnetic ferrites in the suspension of the lumen fibers and stirring rapidly, the internally porous paper fibers are filled with magnetic particles ([14], [15]). In this case, the magnetic functionality is embedded within the body of the fibers, and acts on the entire paper volume. Due to the presence of magnetic contaminants during the cellulose synthesis process, the interconnection between fibers is weakened. It was reported that these magnetic papers exhibit superparamagnetic behavior [16].

---

[1] These prices were obtained through quotations of local suppliers in Germany.

Based on the classical design of spintronics, where magnetic layers are deposited onto inorganic substrates and patterned by means of thin film technology to meanders and concentrators, we would like to realize this concept onto paper substrates. In this case, the magnetic functionality is mainly affected by the surface topology of paper unlike the composite paper embedded with magnetic particles, where internal porosity is more crucial. In addition to the internal porosity of paper fibers [17], conventional paper exhibits large surface roughness due to the anisotropic network topology of the fibers (inter-roughness). Besides, each micro-fibril exhibits nano-roughness and nano-porosity at the surface level (intra-roughness and intra-porosity) (Figure 1).

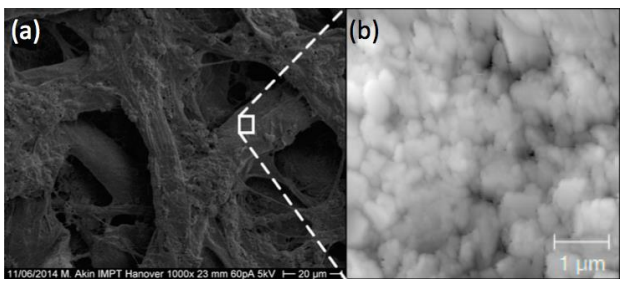

**Figure 1** Example of a topology of paper obtained by scanning electron microscopy (a) and atomic force microscopy (b).

In the following, we review various magnetic phenomena that could occur due to the aforementioned geometrical properties of paper surfaces. In fiber network topologies, domain walls form where fibers/wires meet or change orientation, which can be clearly illustrated on the basis of a semi-circular wire [20]. The presence of domain walls induce scattering of electrons, and therefore electrical resistance [21]. It was also identified that rough surfaces with a network topology allow for domain wall motion easier than rough surfaces without a network topology [18]. In studying artificial spin ice networks, it was observed that each link with a distinct orientation to the magnetic field exhibits locally anisotropic magneto-resistance [23]. Hence, the response of the network is an accumulation of the various local magneto-resistive responses.

Generally speaking, ferromagnetic layers once deposited onto rough surfaces are not expected to exhibit an elemental magneto-resistive response. In this regard, even if not explicitly applied to paper substrates, the impact of surface roughness and surface topology of the substrate on the magnetic behavior of thin layers was thoroughly studied. In the low surface roughness regime (< 10 nm), it was identified that layers deposited on rough surfaces are less sensitive to magnetic fields, which is given by a lower change in magneto-resistance [22]. For intermediate surface roughness regimes (< 100 nm), it was observed that the magnetization reversal depends heavily on the surface roughness, such that the magnetic domain walls once pinned onto rough surfaces can hardly nucleate and propagate, especially with the increase in the ratio of mean roughness to film thickness ([18], [19]). Hence, the ferromagnetic layer becomes more resistant to the magnetic field having a large magnetic coercivity, which is disadvantageous for sensing applications.

Due to the shape anisotropy of every single fiber, the easy axis of magnetization naturally forms along the length of the fiber. Hence, every fiber forms an independent magneto-resistive element contributing to the overall magneto-resistive response of the magnetic paper system. Besides, electrical current is also expected to flow through the length of a fiber. Each fiber becomes therefore an independent electrical conductor with an induced magnetic field. Hence, the direction of easy axis of magnetization and electrical current in a fiber system are expected to be parallel. We speak of nominal electrical current when we consider regions consisting of two or more fibers. Last, the nano-pores of the paper fibrils may form a magnetic tunnel junction or a tunnel magneto-resistance with the surrounding ferromagnetic layer [24]. Similar to the tunneling principle applied in scanning tunnel microscopy [25], tunneling of electrons may occur through the air that is enclosed in the nano-pores.

**Experimental**

We studied the sensitivity of permalloy (Py:$Ni_{81}Fe_{19}$) to magnetic loading when deposited on clean room paper. In order to draw reliable conclusions, we compared the performance of the paper-based system to: (1) Py:$Ni_{81}Fe_{19}$ on polished fused silica ($SiO_2$) and (2) Py:$Ni_{81}Fe_{19}$ on a replication of the paper surface onto an epoxy resin that is mechanically more rigid than paper.

**Materials**

We used clean room paper that has a grammage of 80 g/m², a thickness of 79 μm and a surface roughness given by the root mean squared $R_{rms}$ = 2.877 μm according to the measurement standards DIN EN ISO 4288:1998 (Figure 2). Besides, the paper was impregnated with latex for clean room usage purposes. By computational analysis of the scanning electron microscope image of the paper surface, the surface porosity was determined to be in the range of 23% (Figure 3).

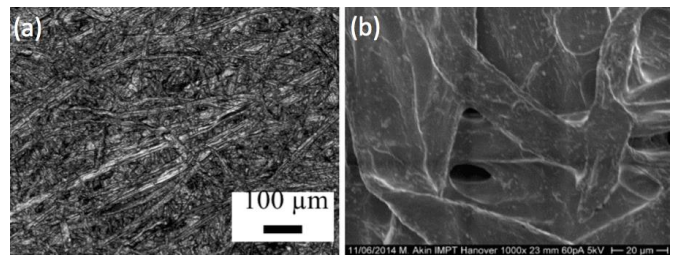

**Figure 2** Surface topology of clean room paper obtained by confocal laser microscopy (a) and scanning electron microscopy (b).

We replicated the surface topology of paper by means of vacuum casting. First, we created a silicone mold of the paper surface. Then, we replicated the surface of the silicone mold onto an epoxy resin (Figure 4 and Figure 5). Due to the effect of gravity during casting, it should be noted that some of the surface porosity could not be transferred from paper to epoxy. Instead, some pores in the paper were transformed to roughness asperities on the epoxy surface. Also, the edges of the epoxy samples were not perfectly straight because they were allowed to remain as free surfaces during the curing process. Similarly, mechanical dicing of fused silica substrates induced chipped edges. Besides, the paper master used for replication could not be used for an actual system since it was contaminated and pre-stressed. In terms of mechanical hardness, paper has a hardness of 189.57 MPa, which was obtained by means of nano-indentation experiments (Berkovich tip). In comparison, the epoxy resin has a D1 shore hardness of 85, which translates to approximately 688 MPa in nano-hardness [30]. Therefore, the paper and epoxy resin systems were expected to respond similarly but not identically. Yet, the replication was expected to be sufficient to draw some first conclusions regarding the impact of the surface topology on magneto-resistance.

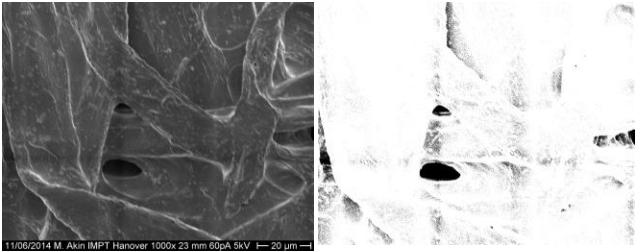

**Figure 3** Image recognition of pores and air gaps from SEM image.

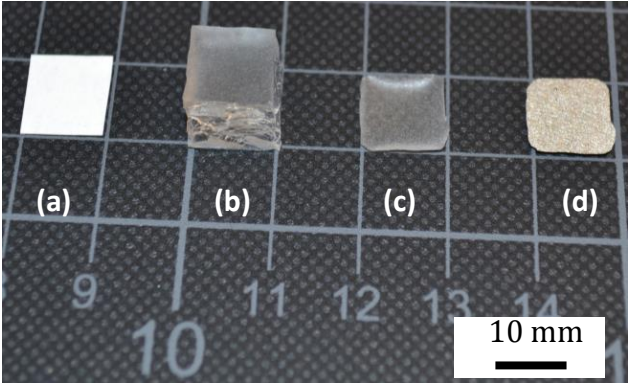

**Figure 4** Replication of paper surface topology to an epoxy resin. (a) Original paper material (b) silicone mold after casting (c) epoxy replica of silicone mold (d) epoxy sample coated with Py:$Ni_{81}Fe_{19}$.

Last, we employed amorphous fused silica ($SiO_2$) with high surface smoothness as reference surfaces. The fused silica glass has a high melting temperature (1710°C), large electrical resistance ($10^{18}$ Ωcm at room temperature) and a mohs hardness of 6, which corresponds to a shore hardness of 89. We opted for a pure $SiO_2$ substrate instead of thermally grown $SiO_2$ on silicon in order to avoid any shunt resistances and the formation of shottky diodes between silicon and Py:$Ni_{81}Fe_{19}$ during the four-point measurements.

By means of a low-temperature solvent-free coating process -sputter deposition-, we coated the three material platforms with Py:$Ni_{81}Fe_{19}$. Decisive for the mechanical quality of the coated systems are the differences in coefficients of thermal expansion between the various materials. In this regard, the coefficients of thermal expansion of the paper and Py:$Ni_{81}Fe_{19}$ were measured using an optical dilatometer (Table 1). It is to be noted that the sputter deposition process was conducted simplistically. Employing electron dispersive x-ray spectroscopy, we confirmed composition homogeneity of the magnetic coating (81 w.% Ni, 19 w.% Fe) on each substrate and identical composition of the coating on all substrates, and thus inertness of all substrate materials to Py:$Ni_{81}Fe_{19}$ (Figure 6).

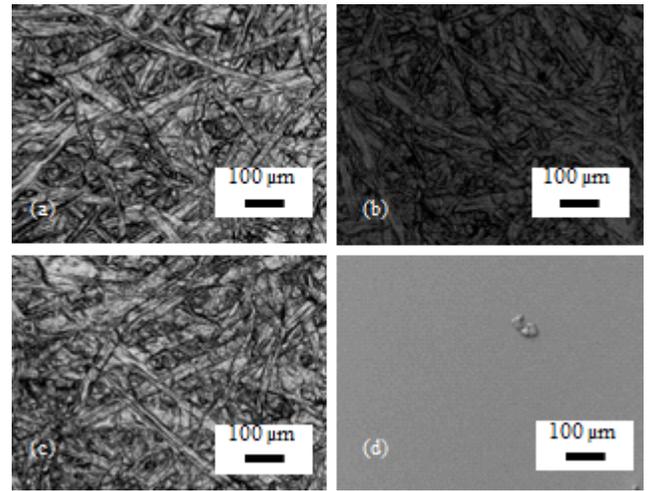

**Figure 5** Laser scanning micrographs of the surface topology of paper, $R_{rms}$ = 2.877 µm (a), silicone cast (b), epoxy replica, $R_{rms}$ = 3.235 µm (c) and fused silica glass $R_{rms}$ = 0.01 µm (d).
(Please note that micrographs (a), (b) and (c) were not taken at consistent locations).

|  | **Fused Silica** | **Clean room paper** | **Py:$Ni_{81}Fe_{19}$** |
|---|---|---|---|
| **20°C** | 0.51 | 210 | -12 |
| **50°C** | 0.51 | -921 | 16 |
| **100°C** | 0.51 | -2816 | 63 |
| **150°C** | 0.58 | -4701 | 110 |

**Table 1** Coefficients of thermal expansion of paper and Py:$Ni_{81}Fe_{19}$ obtained from optical dilatometer measurements and those of fused silica as reported by the material supplier (in [ppm/K]). [2]

Using a vibrating sample magnetometer (VSM), we conducted a comparative study about the dependence of the magnetic properties of Py:$Ni_{81}Fe_{19}$ on the various substrates that we employed in this work. We prepared 5x5 mm² samples according to the preparation guidelines as described in detail in the next section. We measured three samples of each material combination (Py:$Ni_{81}Fe_{19}$ on fused silica, Py:$Ni_{81}Fe_{19}$ on epoxy replication of paper and Py:$Ni_{81}Fe_{19}$ on paper) at 21°C and 40-70 % relative humidity, and commented on the averaged result.

---

[2] Remarkable are the reproducible negative coefficients of thermal expansion of the clean room paper.

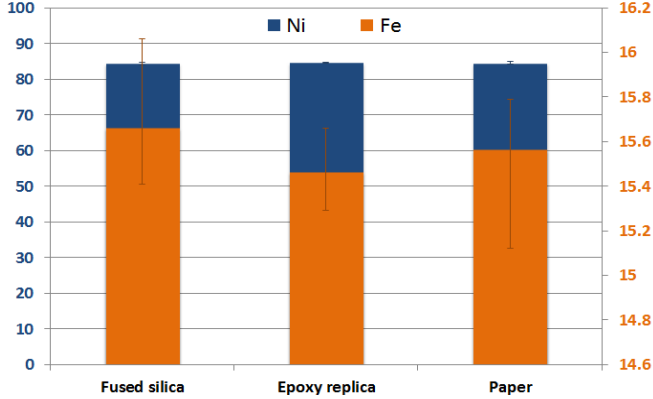

**Figure 6** Electron dispersive x-ray spectroscopy of the Py:$Ni_{81}Fe_{19}$ coating on all three substrate materials.

As expected, we observed an increase in coercivity and coercivity squareness and a decrease in remanent magnetization with the increase in the entropy of the surface topology (Figure 7, Figure 8). While the paper and epoxy replica exhibited similar magnetic properties, paper systems exhibited 65% more coercivity and 35% less remanence than the fused silica systems.

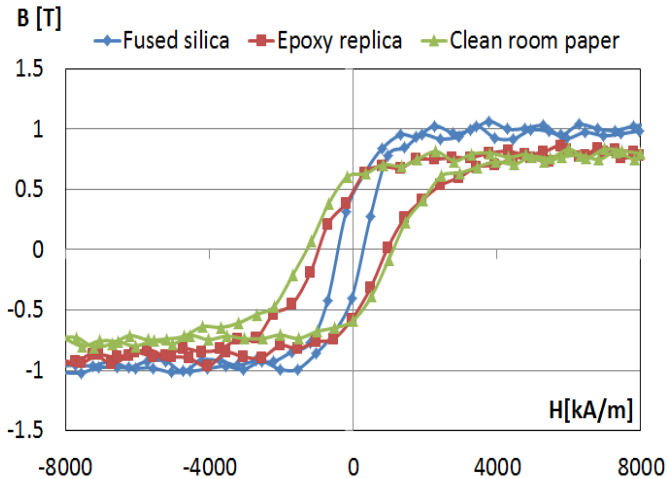

**Figure 7** Magnetic hysteresis (magnetic flux density B vs. magnetic field strength H) of Py:$Ni_{81}Fe_{19}$ as deposited onto fused silica, epoxy replica of paper and paper.

It should be noted that the theoretical coercivity and remanence values of bulk Py:$Ni_{81}Fe_{19}$ were not obtained due to a combination of the following: (1) contamination during the sputter process with other materials, (2) scratches due to handling with tweezers, (3) non-optimized sputtering process with regard to thickness, intrinsic stress and grain structure.

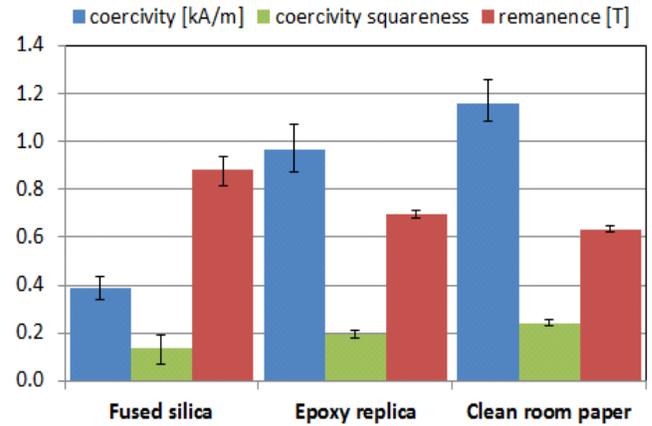

**Figure 8** Coercivity, coercivity squareness and remanence of Py:$Ni_{81}Fe_{19}$ as deposited onto fused silica, epoxy replica of paper and paper.

**Sample preparation guidelines**

In order to exclusively study the effect of surface topology on the magneto-resistance of the clean room paper, we omitted or alleviated as much as possible the occurrence of interfering phenomena due to sample preparation. First, we prepared square-shaped samples of the size of 10x10 mm², decreasing the effect of surface anisotropy on the magneto-resistance measurements.

It was observed that machining of substrates after deposition of magnetic layers created boundary defects and geometrical singularities in the magnetic layer, which affect the magnetic performance of the system [26]. Since we employed materials that were machined with different techniques (dicing for fused silica and cutting for paper), we machined the substrates prior to sputter deposition. The epoxy samples were fabricated in required size and did not go through any additional machining after coating.

Next, we minimized the occurrence of large mechanical stresses within the Py:$Ni_{81}Fe_{19}$ coating due to mismatches in thermal expansion when cooling down after the sputter deposition process (Table 1). Besides, the paper and epoxy substrates were expected to release layer stress easier than fused silica due to mechanical interlocking between Py:$Ni_{81}Fe_{19}$ and the substrates. In order to eliminate preferential stress formation, we employed a low-power sputtering process (50 W) such that the substrate and layer temperature remained below 50°C.

All samples (four of each material configuration) were prepared within one single sputter deposition process. In this case, we removed any fluctuations in layer thickness and chamber conditions from batch to batch. In order to preserve the material surface topology, we deposited Py:$Ni_{81}Fe_{19}$ layers of a thickness of 70 nm (<$R_{rms}$ of the rough substrates). The texture of the substrate surface and the stochastic deposition pattern of the Py:$Ni_{81}Fe_{19}$ particles during sputter deposition determine the initial energy state of the coating. In order for the Py:$Ni_{81}Fe_{19}$ coating to reach its minimal energy state before magneto-resistive measurements, the samples are pre-conditioned by subjecting them to magnetic field cycling.

**Measurement set-up**

We used a four-point-measurement setup built within a concentrated magnetic field of the size of 18.8 x 20.45 x 29.44 mm$^3$ (Figure 9). By inducing an electric current into wire wound coils around a soft magnetic metal, the magnetic field was created (Figure 9). A maximum magnetic intensity of 90 kA/m (~1100 Oe) could be applied with this setup. In order to explore the magneto-resistivity of Py:$Ni_{81}Fe_{19}$, a magnetic intensity of 8 kA/m (~100 Oe) is known to be more than sufficient to reach a magnetic saturation state. The magnetic field was applied at the four-point-probe location. Each measurement cycle started at -8 kA/m. At discrete steps of 0.025 kA/m, the magnetic intensity was swept to 8 kA/m and back to -8 kA/m. Each discrete magnetic intensity was held constant for 1 ms before sweeping continues to allow for transient transition between two discrete magnetic intensities.

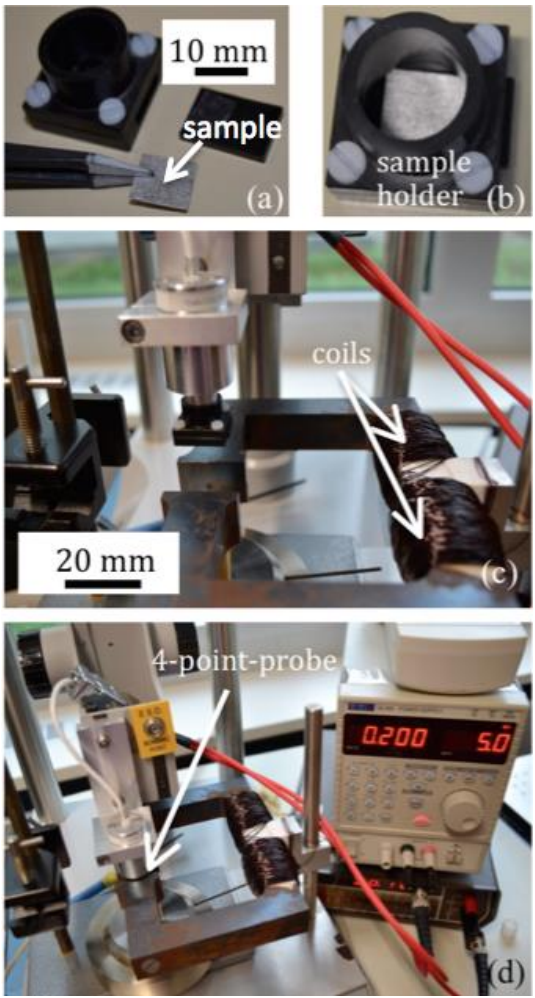

**Figure 9** (a) Handling of sample, (b) sample holder to prohibit paper from bending, (c) descending of four-point-probe into magnetic field, (d) induction of constant electric current into coils and creation of magnetic field.

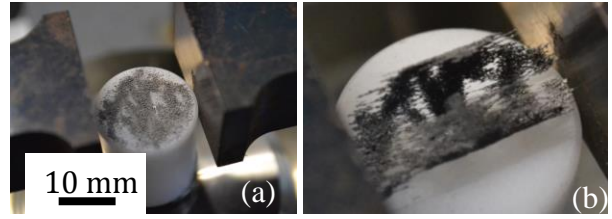

**Figure 10** (a) Iron particles before application of magnetic field, (b) visualization of magnetic field with iron particles at maximum magnetic intensity of the measurement setup at the 4-point-probe measurement location.

Since Py:$Ni_{81}Fe_{19}$ conventionally manifests anisotropic magneto-resistance, we measured the resistance of Py:$Ni_{81}Fe_{19}$ in two electric current-to-magnetic-field configurations. First, we measured the parallel configuration: The nominal flow orientation of the electric current applied to Py:$Ni_{81}Fe_{19}$ is parallel to the magnetic field lines. Then, we measured the perpendicular configuration: The nominal flow orientation of the current applied to Py:$Ni_{81}Fe_{19}$ is perpendicular to the magnetic field lines. It is to be noted that the current flow at the measurement location could be directed. Outside of the measurement location, the direction of the current flow could not be preset due to lack of a patterned coating.

During the measurements, the ambient temperature and relative humidity were regulated at 21°C and 40-70 % respectively. Besides, we handled all samples using tweezers and ESD treatment measures in order to not induce any thermal or electrical perturbation to the system.

We measured the relative magneto-resistance coefficient as given by:

$$MR_r [\%] = 100 \cdot (R_c - R_{min})/R_{min}, \quad (1)$$

where $R_{min}$ is the minimum electrical resistance recorded through the measurement cycle and $R_c$ is the instantaneous electrical resistance.

**Noise analysis and measurement errors**

While considering our measurement results, we would like to point out sources of noise during measurement that we were not able to control despite extreme caution. First, the co-centric positioning of the sample to the 4-point-probe was done manually. Hence, the position of the four-point-probe slightly deviated from the center of the sample with small, yet different, amounts at each experiment. Besides, a remaining magnetization of 1.7 kA/m was present in the magnetic region even when the measurement setup was not in operation. Thus, samples when positioned into the measurement setup were first subject to the remaining magnetic field, which is alleviated with the pre-conditioning step.

The four-point-probe disposes of rounded needles with a radius of 500 μm that are individually biased by a spring load of 1-2 N/mm$^2$. After measurement, we observed a remaining indented area in the soft systems like paper and epoxy (Figure 12). A cracking of Py:$Ni_{81}Fe_{19}$ due to indentation and an increase in the primary electrical resistance of Py:$Ni_{81}Fe_{19}$ around the gauss probes were therefore expected.

In order to estimate the order of magnitude of the regular shunt resistances occurring in the fused silica–based system, we measured the electrical resistance of the bare fused silica substrates. Let $R_{Py:Ni81Fe19}$, $R_{SiO2}$ and $R_{total}$ be the electrical resistances of the Py:$Ni_{81}Fe_{19}$ coating, the $SiO_2$ substrate and the total of the parallel resistances Py:$Ni_{81}Fe_{19}$ and $SiO_2$ (($R_{SiO2}$ + $R_{Py:Ni81Fe19}$)/( $R_{SiO2}$ . $R_{Py:Ni81Fe19}$)). In the case of polished $SiO_2$ substrates, we determined a ratio of ($R_{total}$ / $R_{SiO2}$) = 0.44. For rough $SiO_2$ substrates, we obtained a ratio of ($R_{total}$ / $R_{SiO2}$) = 0.02. Consequently, we deduce that the polished $SiO_2$ systems are more susceptible to shunt resistances. During the DC four-point measurements, we have also observed a temporal drift due to thermal noise that was more remarkable with the paper and epoxy resin systems. We have filtered this noise from the measurements up to a fourth order polynomial approximation.

In addition, we explored the systematic noise originating from the set-up in the absence of all inputs (sample, electrical current and operated magnetic field). By applying the four-point probe to the insulating bottom support to the samples, used during actual measurement, we track the fluctuations $V_i$ in n voltage measurements by the four-point-probe, where n = 70. In both measurement configurations, we recorded average deviations of (($\Sigma V_i^2$)/n)$^{1/2}$ = 0.64 mV, which translate to a 6.24 %-deviation for fused silica systems, up to a 2.2 %-deviation for the epoxy-based systems and a maximum deviation of 1.12 % for the paper-based systems.

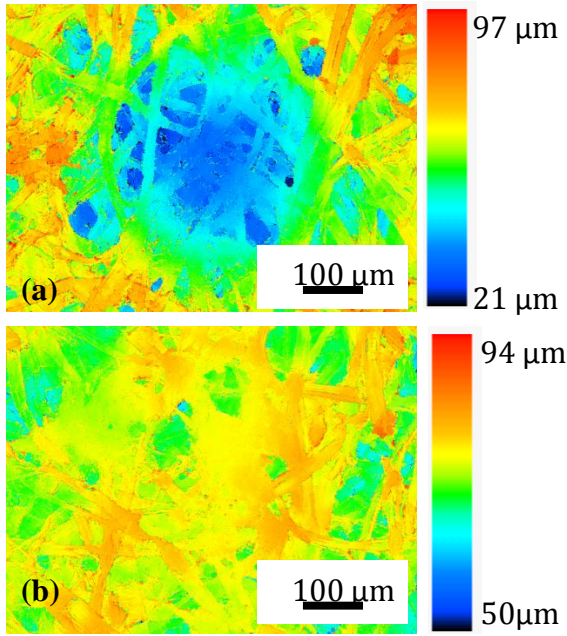

**Figure 11** Remaining indentation on (a) paper surface and (b) epoxy surface after four-point-measurement.

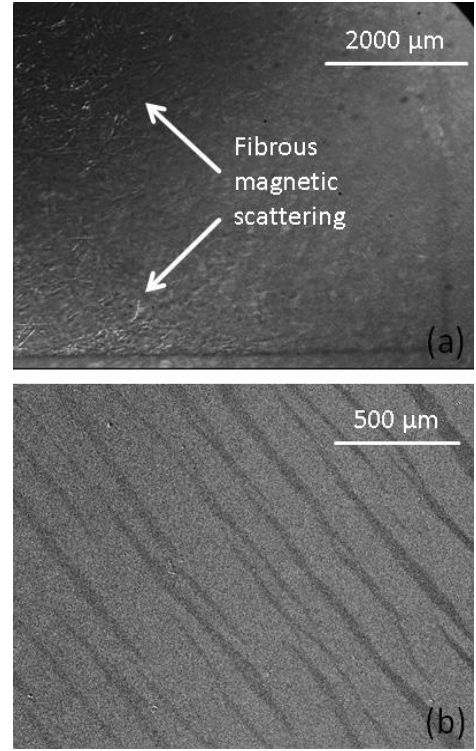

**Figure 12** Magneto-optic kerr effect micrographs of (a) Py:$Ni_{81}Fe_{19}$ on paper and (b) Py:$Ni_{81}Fe_{19}$ on fused silica.

**Results and discussion**

Qualitatively speaking, all three systems exhibited a change in magneto-resistance. As expected, the fused silica systems exhibited a decrease in magneto-resistance due to the planar Hall effect in the parallel configuration, and an increase in magneto-resistance at the perpendicular configuration ([27], [29]). When comparing the magneto-resistive behavior of paper to the epoxy replica, we observed similar phenomenological behavior, most probably due to the macroscopic similarity of the surface topology (Figure 13, Figure 14). As described above, each fiber constitutes a distinct electrical conductor and magneto-resistor, with parallel direction of easy axis of magnetization and electric current flow. Due to the stochastic orientation of the paper fibers on the surface, the superposition of the individual fiber responses averages closely to a classical ±45 ° response on highly smooth substrates for both measurement configurations. The latter behavior was similarly observed in artificial spin ice networks 23]. Using the magneto-optic kerr effect, we were able to locate the formation of domain walls at the bending location of fibers (Figure 12). However, we did not observe any direct implication of the expected tunneling of electron waves through the intra-porosity and -roughness of the paper. Since the routing of the electrical current in the non-patterned Py:$Ni_{81}Fe_{19}$ coating was not fully controlled, the percentages of change in magneto-resistance were not maximized.

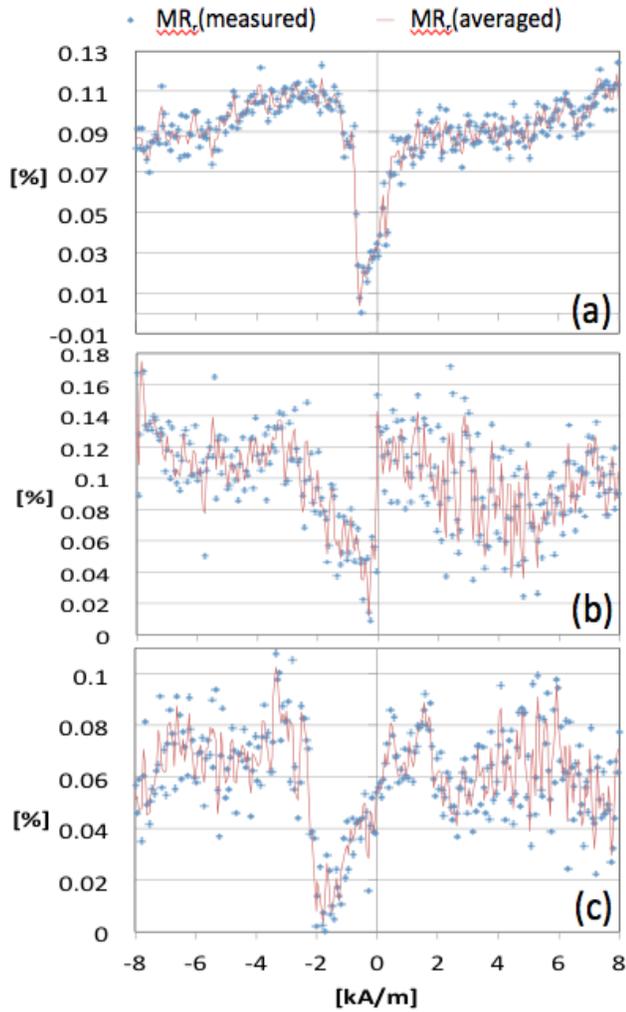

**Figure 13** Evolution of coefficient of magneto-resistance $MR_r$ [%] in the parallel configuration for the (a) Py:$Ni_{81}Fe_{19}$-on-fused silica glass, (b) Py:$Ni_{81}Fe_{19}$-on-replication of paper on epoxy resin and (c) Py:$Ni_{81}Fe_{19}$-on-paper systems.

## Conclusions and outlook

In this work, we assessed the magneto-resistive behavior of Py:$Ni_{81}Fe_{19}$ deposited onto paper substrates. As reference systems, we employed Py:$Ni_{81}Fe_{19}$ deposited onto fused silica and onto a replication of the paper surface in an epoxy resin. Qualitatively speaking, all three systems exhibited an anisotropic change in magneto-resistance. Due to the stochastic orientation of the paper fibers on the surface, the superposition of the individual fiber responses averages closely to a classical ±45° response onto highly smooth substrates for the measurement configurations: (1) nominal electrical current and magnetic field are parallel, and (2) nominal electrical current and magnetic field are perpendicular. Using the magneto-optic kerr effect, we have identified the bending locations of the fibers as domain walls. We argue that these paper-based systems will find interesting applications, especially due to the magneto-resistive sensitivity that we have observed in this work.

Based on these promising results, we plan to further investigate the physics behind the magneto-resistive sensitivity of Py:$Ni_{81}Fe_{19}$ deposited onto paper substrates. In particular, we would like to investigate the responsivity and reproducibility of the Py:$Ni_{81}Fe_{19}$-on-paper systems at different current-to-magnetization angles. Besides, we investigate the performance of these systems in dependence of the paper geometry (micro- and nano-papers), paper flexibility, thickness and structure of the Py:$Ni_{81}Fe_{19}$. Once the interplay between Py:$Ni_{81}Fe_{19}$ and paper is well understood, the intrinsic magnetic properties of the sputter deposited Py:$Ni_{81}Fe_{19}$ will be further optimized to achieve high performing systems. For instance, annealing techniques and sputter process parametrization will be researched.

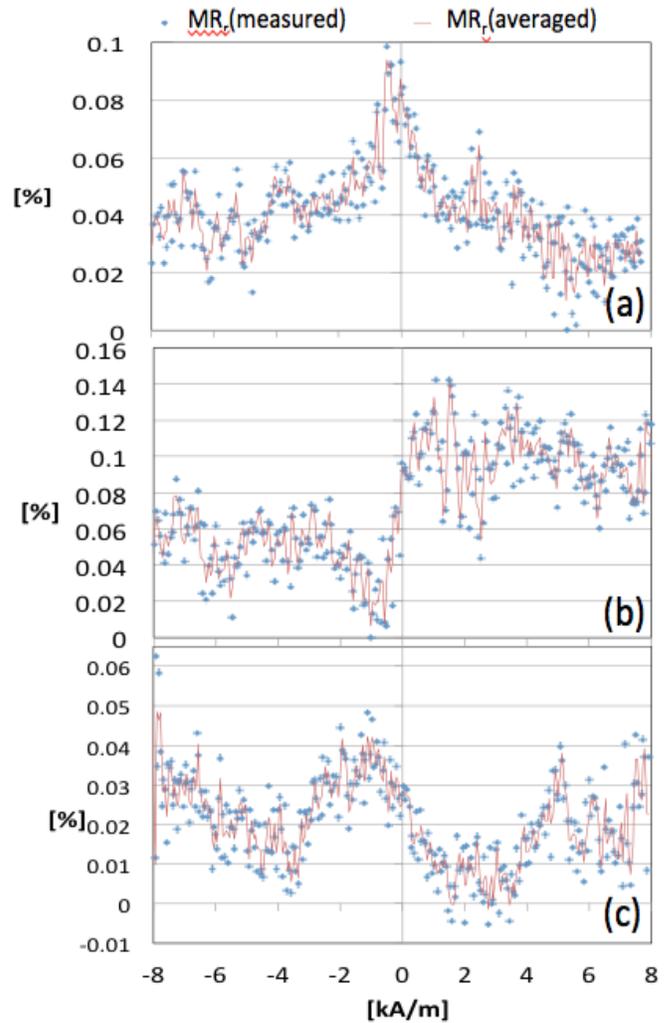

**Figure 14** Evolution of coefficient of magneto-resistance $MR_r$ [%] in the perpendicular configuration for the (a) Py:$Ni_{81}Fe_{19}$-on-fused silica glass, (b) Py:$Ni_{81}Fe_{19}$-on-replication of paper on epoxy resin and (c) Py:$Ni_{81}Fe_{19}$-on-paper systems.

## Acknowledgments

This work was partially funded by the German Research Foundation. We would like to thank Elke Pichler from the Energy Research Center of Lower Saxony at Technical

University of Clausthal for conducting the optical dilatometer measurements. Also, we would like to thank Tanja Marjov from the Institute of Micro Production Technology for conducting the nano-indentation experiment on paper. Further, we would like to acknowledge Rudolf Schaefer and Stefan Pofahl from the Leibniz Institute of Solid State and Materials Research Dresden for their assistance with the magneto-optical kerr effect microscopy measurements. Last, we would like to thank Piriya Taptimthong, Rahel Kruppe, Anja Wienecke, Lisa Jogschies and Johannes Rittinger from the Institute of Micro Production Technology and Sami Akin from the Institute of Communications Technology at Leibniz Universitaet Hannover for helpful discussions.